\documentclass[10pt,journal,cspaper,compsoc]{IEEEtran}
\ifCLASSOPTIONcompsoc
\else
\fi
\usepackage[pdftex]{graphicx}
\usepackage[pdftex]{hyperref}

\usepackage{mathtools}
\usepackage{stmaryrd}
\usepackage{multirow}

\ifCLASSINFOpdf
\else
\fi
\newcommand{\heading}[1]{\multicolumn{1}{|c|}{\bfseries #1}}

\begin{document}
%
\title{{Performance-oriented Cloud Provisioning: Taxonomy and Survey
\thanks{This work is submitted to a journal for possible publication. Copyright may be transferred without notice, after which an accepted version will be uploaded and accessible. A small portion of this work has been published earlier in the \emph{Proceedings of the 2012 IEEE/ACM Fifth International Conference on Utility and Cloud Computing (UCC 2012)}.}
}}
%
%
%
%
%

\author{Yasir~Shoaib
        and~Olivia~Das
\IEEEcompsocitemizethanks{\IEEEcompsocthanksitem Y. Shoaib and O. Das are with the Department
of Electrical and Computer Engineering, Ryerson University, Toronto,
ON, Canada, M5B 2K3.\protect\\
E-mail: yasir.shoaib@ryerson.ca, odas@ee.ryerson.ca}
\thanks{}}

%
%

\markboth{November~2014}%
{Shoaib and Das: Performance-oriented Cloud Provisioning: Taxonomy and Survey}
%


\IEEEcompsoctitleabstractindextext{%
\begin{abstract}
Cloud computing is being viewed as the technology of today and the future. Through this paradigm, the customers gain access to shared computing resources located in remote data centers that are hosted by cloud providers (CP). This technology allows for provisioning of various resources such as virtual machines (VM), physical machines, processors, memory, network, storage and software as per the needs of customers. Application providers (AP), who are customers of the CP, deploy applications on the cloud infrastructure and then these applications are used by the end-users. To meet the fluctuating application workload demands, dynamic provisioning is essential and this article provides a detailed literature survey of dynamic provisioning within cloud systems with focus on application performance. The well-known types of provisioning and the associated problems are clearly and pictorially explained and the provisioning terminology is clarified. A very detailed and general cloud provisioning classification is presented, which views provisioning from different perspectives, aiding in understanding the process inside-out. Cloud dynamic provisioning is explained by considering resources, stakeholders, techniques, technologies, algorithms, problems, goals and more.
\end{abstract}

}

\maketitle

\IEEEdisplaynotcompsoctitleabstractindextext

%
\IEEEpeerreviewmaketitle

\section{Introduction}
\label{sec:Introduction}

\IEEEPARstart{A}{pplication} providers (AP) strive for deploying web applications that meet high performance standards even when workload demands are at their peak. They meet these request demands either through the management of their own web-server farms or through the purchase of hardware and software as services from cloud computing service providers.

Considering time, space, cost, and flexibility as dimensions, the selling point of cloud computing is the ``multi-dimensional ease'' \cite{2012_ucc2012YasDas} with which computing resources could be accessed through the Internet. The resources, viz. processors, network and software remain in cloud provider's data centers and can be added and removed as needed by the customer \cite{NIST_CloudDefn,velte2009cloud1}. With cloud computing's utility-based pricing model, the customers pay for the resources used; in comparison, they would have incurred capital expenses if resources were purchased \cite{velte2009cloud1}. These customers considering their spatial, temporal and monetary constraints, could thereby rely on services the ``cloud'' \cite{zhang2010cloud13} is known to offer. The well-known cloud services offered by providers such as Amazon, Google and Microsoft are: Infrastructure-as-a-Service (IaaS), Platform-as-a-Service (PaaS), and Software-as-a-Service (SaaS); however, the levels of service provided vary from one CP to another \cite{ranjan2010peer12,zhang2010cloud13}. For provisioning networks and physical machines, services such as Network-as-a-Service (NaaS) \cite{2012_NAAS_Costa,2011_DynamicallyScalingApps}, and Metal-as-a-Service (MAAS) \cite{2013_MAAS_Canonical,2012_MAASEffect}, respectively, can also be accessed through the clouds.

Cloud computing, however, faces various technical challenges related to application hosting. One of these deal with autonomously provisioning adequate resources, i.e. dynamically adding and removing cloud resources to handle the fluctuating Internet user request demands of the applications \cite{zhang2010cloud13}. Under-provisioning of resources cause application end-users to experience excessive delays, especially during demand surges. Eventually, due to poor performance, disgruntled users stop using the application, incurring loss to the AP businesses --- as seen happening with eCommerce sites \cite{ForresterWhitepaper34}. On the other hand, over-provisioning leads to higher costs for the cloud provider (CP) due to management of large number of servers that because of being under-utilized cause excess power consumption and heat dissipation in their data centres. The loss is not only restricted to CP but also extends to AP, who have constrained budgets but have to pay for unnecessary VM instances running. The quintessential scenario would be the dynamic provisioning of resources following the fluctuating demands to meet various Quality-of-Service (QoS) requirements and reduction of cost and power, which in practice is quite a challenging endeavor \cite{zhang2010cloud13}.

When resources are allocated once and the system doesn't adjust itself to varying workload requirements, such a provisioning is static in nature. This static provisioning causes either under or over-provisioning of resources \cite{2012_EmpiricalPredicationModels}. In contrast, dynamic provisioning solves the under and over provisioning of resources by adjusting resource allocations to the changing workload that the system receives. Meng et al. \cite{2010_EfficientResourceProvisioning} describe static provisioning and dynamic provisioning for VMs, where the former is generally associated to the first step of capacity planning, which is done monthly or seasonally. To showcase the applicability of their dynamic provisioning approaches, many works have compared their dynamic provisioning approach with the static counterpart (e.g. \cite{2011_DynamicResProvMultiplayerGames,calheirosvirtual3, iqbal2010sla6}). The focus in this article is on dynamic provisioning. Those who employ dynamic provisioning also need an initial deployment of the system, where the numbers of the resources are known. This generally forms an implicit part of dynamic provisioning but the main idea is the adaption of the system to changes.

This article provides a detailed literature survey on dynamic provisioning within cloud systems with focus on application performance. Along with a general and clear classification of cloud dynamic provisioning, this article contributes by explaining the different facets of provisioning in the clouds and in particular explaining in detail the following: reactive and proactive provisioning, horizontal and vertical provisioning, provisioning of resources as services, provisioning goals, algorithms, techniques and technologies involved. 

The structure of the article is as follows: Section~\ref{sec:classifyCloudProv} gives a general introduction to cloud provisioning, providing a detailed and general cloud provisioning classification, which views provisioning from different perspectives, aiding in understanding the process inside-out. Following this, the article explains cloud provisioning by considering decision-making entities, problems, types, scaling, policies, resources, techniques, technologies, algorithms and more. Section~\ref{sec:TypesCloudProvTerm} mentions about the role of cloud provider and application providers in making decisions associated to provisioning. Section~\ref{sec:CloudProvTypesProbPolicies} mentions different provisioning types, scaling, associated problems and the placement policies. Sections~\ref{sec:ReactiveProactiveProv},~\ref{sec:ServiceProv}, and ~\ref{sec:ProvAlgorithms} explain dynamic provisioning based on reactive and proactive provisioning, service-levels, and algorithms. Section~\ref{sec:TowardPerCloudProv} lists works that relate to dynamic provisioning and its progression to provisioning in the clouds with focus on performance. Section~\ref{sec:Conclusions} presents the conclusions.
\begin{figure*}[t]
\centering
\includegraphics[keepaspectratio,trim=14mm 44mm 150mm 27mm,clip,height=15cm]{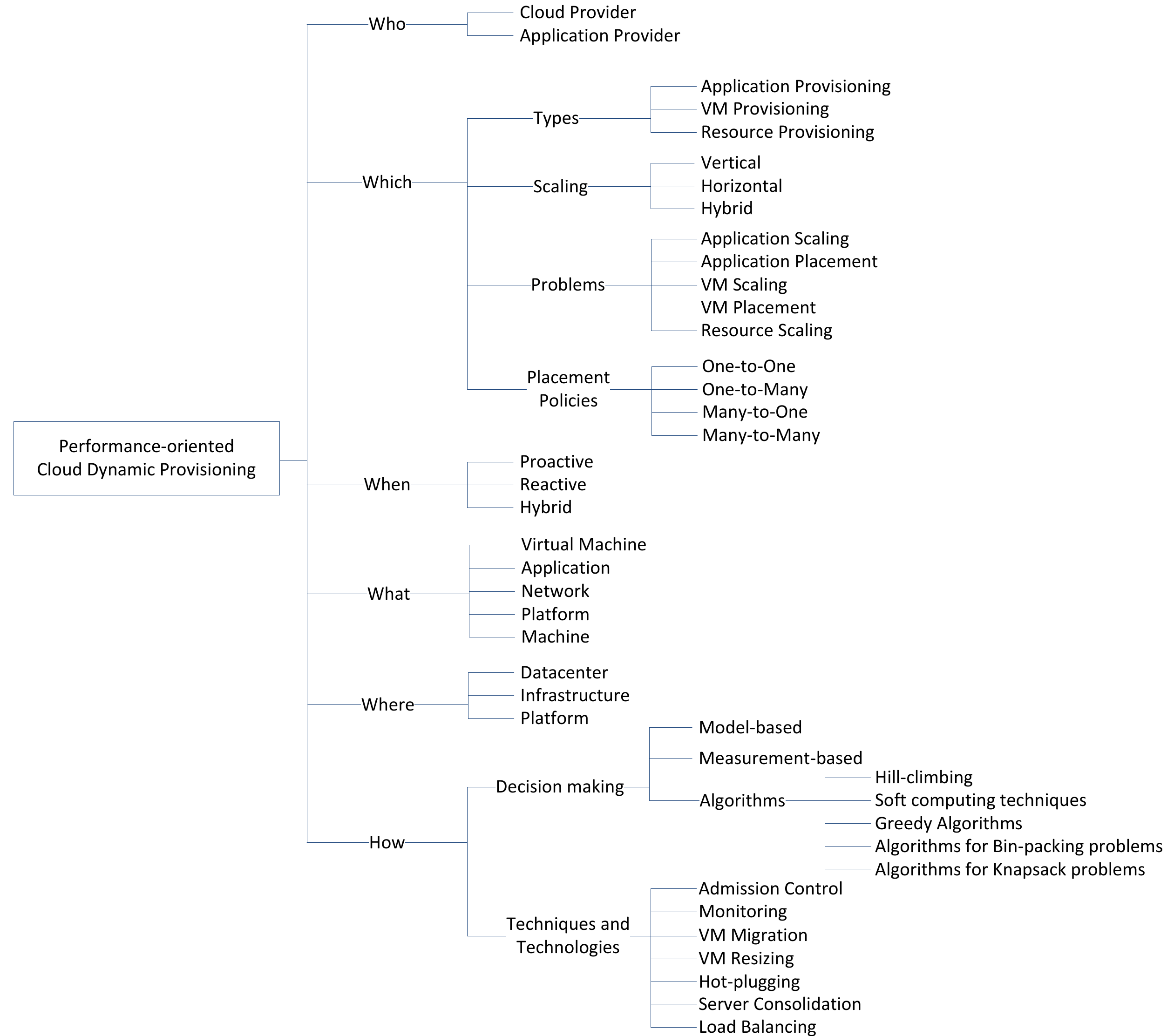}
\caption[Cloud Provisioning - Classification]{Cloud Provisioning - Classification}
\label{fig:CloudProvClassification}
\end{figure*}

\section{Classifying cloud provisioning}
\label{sec:classifyCloudProv}

\emph{Cloud provisioning} is a dynamic provisioning problem applied within a cloud system. Calheiros et al. \cite{calheirosvirtual3} list following as three \emph{steps} --- or as \emph{types} as referred in this article --- to cloud provisioning:
\begin{IEEEenumerate}
	\item \textbf{VM provisioning:} creation of VMs to meet software and hardware based requirements of an application such that given performance levels are achieved.
	\item \textbf{Resource provisioning:} association of created VMs to adequate hardware resources.	
	\item \textbf{Application provisioning:} application deployment in the VMs and the subsequent association of the requests received to those applications.
\end{IEEEenumerate}

There are various ways to classify provisioning in cloud systems. Following is the proposed classification (refer \figurename~\ref{fig:CloudProvClassification}):
\begin{IEEEitemize}
\item \textbf{Who} performs the provisioning (\textit{the decision making entity}). These would be entities that make decisions about when resources need to be added and removed. AP and CP are two such entities. Section~\ref{sec:TypesCloudProvTerm} explains the different decision making entities.
\item \textbf{Which} provisioning \textit{types, scaling, problems and policies} exist. In simple classification, provisioning may be classified as application, VM and resource provisioning. Related to these provisioning types are problems such as application scaling, application placement, VM scaling, VM placement and resource scaling. Section~\ref{subsec:ProvProblems} and Section~\ref{subsec:ProvisioningTypes} explain these in detail. Furthermore, resource scaling may be classified either as horizontal or vertical, where the former type of scaling is based on adding or removing of new or existing resources (e.g. VMs) and the latter is associated with adding of resources to the existing live VMs and machines. Section~\ref{sec:HorizVertScaling} discusses about horizontal and vertical scaling. The relationships between resources and placement policies are discussed in Section~\ref{subsec:PlacPolicies}. One-to-one, one-to-many, many-to-one, and many-to-many policies could be used to define different placements of VMs on physical machines and of applications on VMs.
\item \textbf{When} is the provisioning decision made (\textit{reactive versus proactive provisioning}). This could be after the workload and demands have increased/decreased or before the occurrence of such events. Reactive provisioning suggests provisioning decisions that are made after changes in workload behaviour is noticed whereas proactive provisioning involves prediction based approaches that help prepare ahead of changes in the workload and system usage. However, a hybrid approach may also be applied for building a more robust system, using both reactive and proactive provisioning. Section~\ref{sec:ReactiveProactiveProv} explains about these approaches.
\item \textbf{What} \textit{resource} is being provisioned and allocated, e.g. resources such as applications, VMs, processors, memory, network, storage, etc. Here, applications are allocated on VMs and VMs are allocated on physical machines. Also, there is inter-dependence between resources, e.g. in some cases increasing of processors may be achieved only by instantiation of a new VM and increase in VMs would increase all the resources that the new VM requires for execution. Sections~\ref{sec:ServiceProv} and ~\ref{sec:TowardPerCloudProv} discuss the provisioning of resources as services.
\item \textbf{Where} is the decision-making taking place (\textit{decision-making layer}). CP that provide infrastructure would make decisions on a global-level to optimize resource usage in data centers, whereas AP as IaaS customers would employ localized decisions for each of their applications that run on VMs based on the workload they receive and cost of using the services. At the platform-level the PaaS providers would scale application containers and database systems \cite{2011_DynamicallyScalingApps}. Section~\ref{sec:ServiceProv} and Section~\ref{sec:TowardPerCloudProv} mention various works that make decisions at different service-levels.
\item \textbf{How} is the provisioning problem being solved, relating to the various \textit{algorithms, techniques and technologies} that could be employed for deriving solutions. Commonly used algorithms include hill-climbing, soft computing techniques (e.g. genetic algorithms, neural networks, fuzzy logic etc.), greedy algorithms, bin-packing based algorithms, and knapsack based algorithms, etc. The main techniques and technologies that are applied and form a part of the algorithms are hot-plugging, VM migration, VM resizing, performance modeling, workload prediction, admission control, system monitoring, server consolidation and load balancing. Section~\ref{sec:ProvAlgorithms} list various provisioning algorithms.
\end{IEEEitemize}

The detailed classification is useful in understanding of cloud dynamic provisioning. In the following sections, we first discuss the types, problems and policies associated with provisioning, and then mention works in cloud provisioning.

\section{Who does cloud provisioning?}
\label{sec:TypesCloudProvTerm}

\begin{figure}[h!tb]
\centering
\includegraphics[keepaspectratio,viewport=90 152 630 397,clip,width=0.5\textwidth]{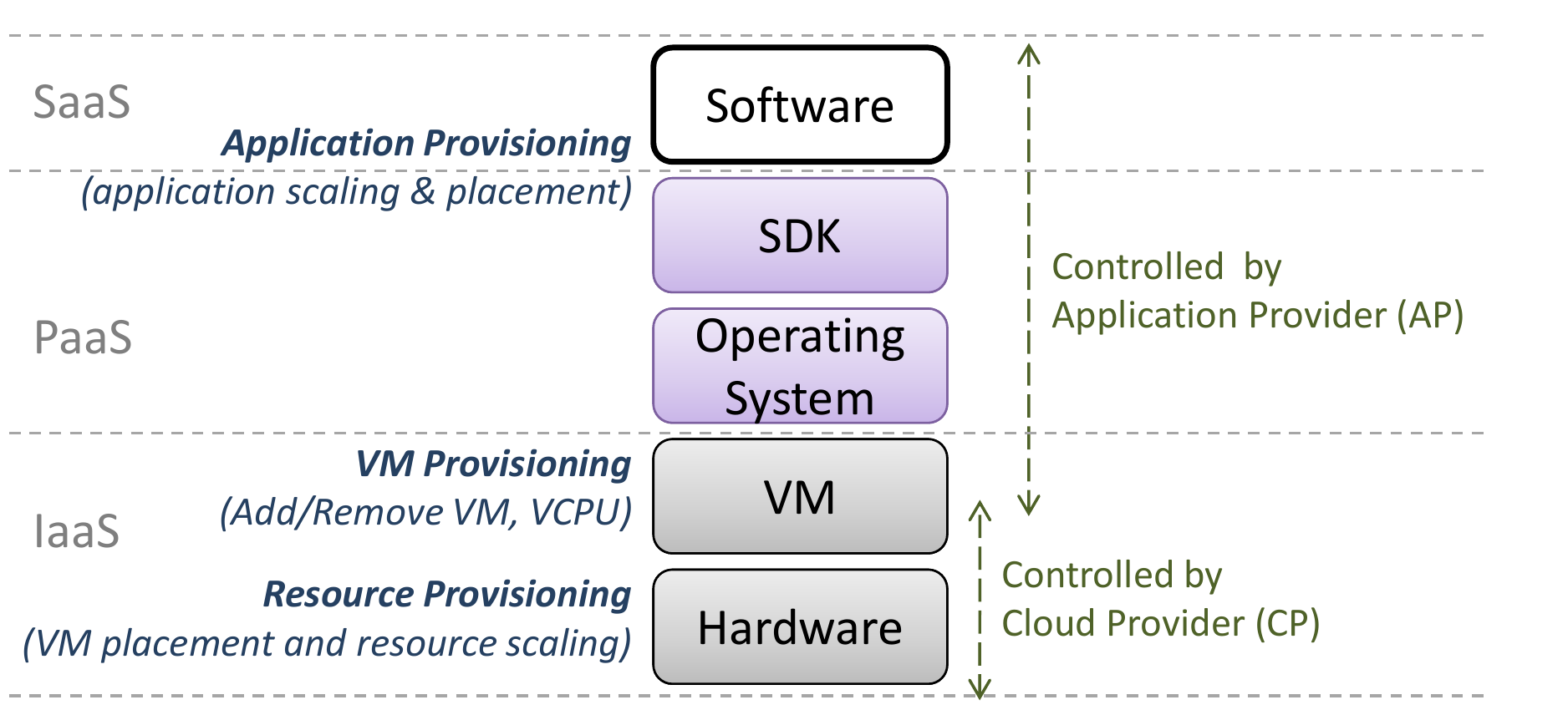}
\caption[Cloud control hierarchy]{Cloud control hierarchy displaying layers controlled directly by CP and AP. CP chooses the hardware to allocate VMs on (resource provisioning), whereas, AP controls: 1) application scaling and placement (application provisioning), and 2) the addition/removal of VM instances and associated resources (VM provisioning). Although, CP houses the resources and as such has control over all layers, it does not directly administer services purchased by the customers. In the paper by Calheiros et al. \cite{calheirosvirtual3}, VM and application provisioning are performed by AP and the authors mention that resource provisioning decisions are made by IaaS provider (i.e. CP). Similar idea of ``decoupled control'' \cite{2009_AutomatedControlCloudChallenges} has previously been suggested by Lim et al. This image is reprinted and presented here with modifications to the image in \emph{Proceedings of the 2012 IEEE/ACM Fifth International Conference on Utility and Cloud Computing (UCC 2012)} \cite{2012_ucc2012YasDas} \copyright\ 2012 IEEE}
\label{fig:cloudControlHierarchy}
\end{figure}

In the cloud provisioning context, we distinguish between CP, AP and end-users \cite{jimliwoodside2009performance5}. \figurename~\ref{fig:cloudControlHierarchy} shows the cloud control hierarchy. CP provide IaaS, which includes hardware and virtualized resources such as VMs that run on the hardware. Its customers directly access the virtual resources only. AP are generally IaaS customers, who can create and instantiate VMs and install their choice of operating environment, which includes operating systems (OS) and software development kits (SDK). These VMs may be allocated and de-allocated on-demand through facilities --- such as API --- provided by the CP. AP develop their web applications and deploy them on VMs, thereby providing SaaS to their customers: end-users.

Each \emph{application} is composed of running \emph{tasks} (or processes). One or more tasks may be grouped together to form a \emph{tier} of the application. A request that the application receives is handled by the first tier and propagated to the next lower tier if further processing is needed, thereby having the request traverse through to the multiple tiers of the application, from one tier to the next. Once the request has been processed by the tiers, the response is sent back to the upper tiers and finally a response is returned to the end-user that initiated the request. In some works, a \emph{service} may also be used to denote an application or its component, e.g. task.

Provisioning decisions made by AP are for accommodating the workload demands of applications that are accessed by end-users. Although, it is possible that AP be PaaS customers only, instead of being IaaS customers, however, for simplicity we generally assume that AP are IaaS customers with ability to create VMs, assign virtual processors and other resources to VMs, and place application on VMs. Furthermore, CP can also deploy their own software and provide SaaS to end-users, in which case they take on the role of AP as well.

\begin{figure*}[ht]
\centering
\includegraphics[keepaspectratio,trim=13mm 63mm 17mm 22mm,clip,width=\textwidth]{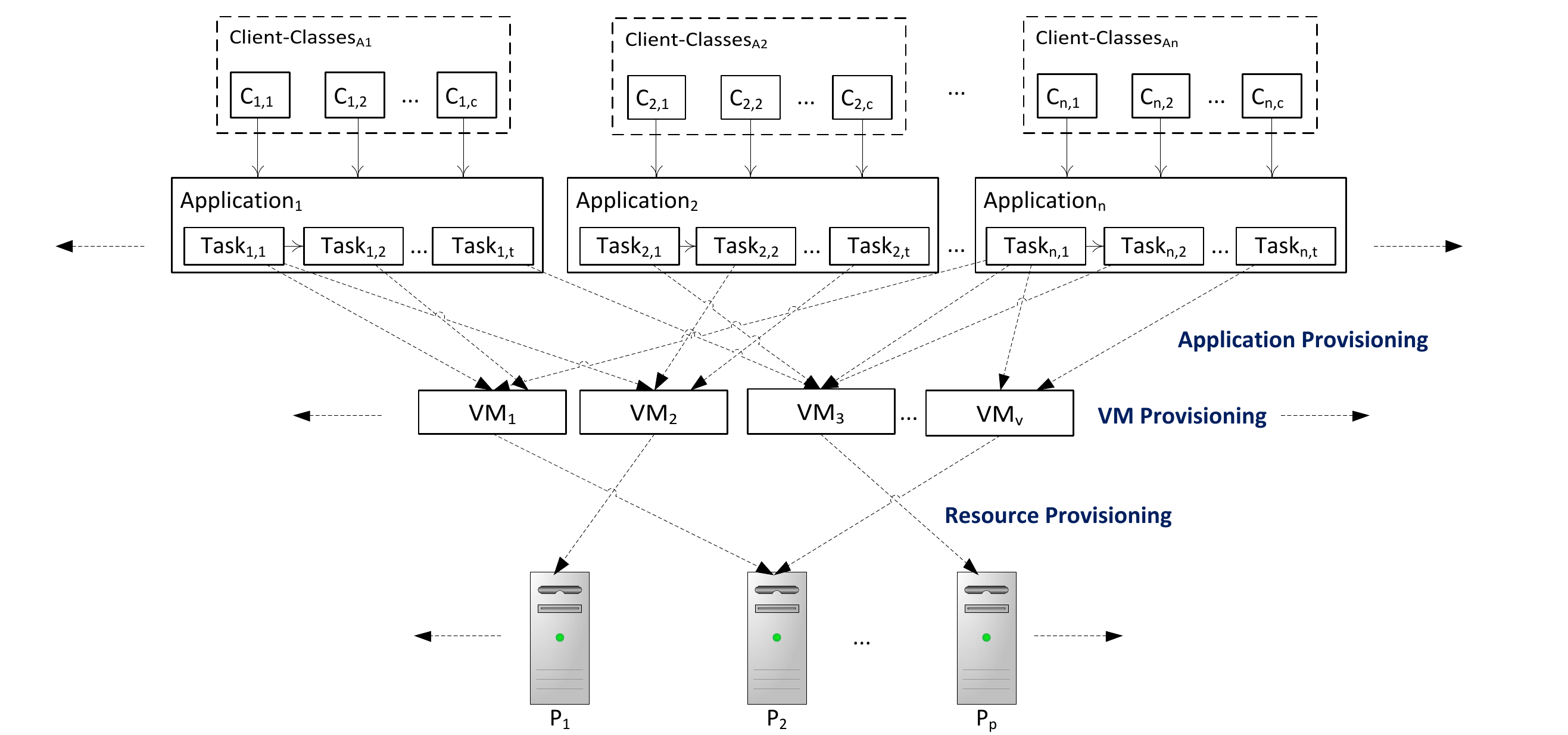}
\caption[Cloud Provisioning]{Cloud provisioning showing applications managed within a cloud. AP are responsible only for application provisioning and VM provisioning. Also refer to ``Dynamic resource allocation'' image that shows a similar idea of  provisioning by Van et al. \cite{2009_AutoVMManagement_NguyenVan}, although more details are present in the image above.}
\label{fig:CloudProv}
\end{figure*}

\section{Which provisioning types, scaling, problems and policies exist?}
\label{sec:CloudProvTypesProbPolicies}
Provisioning of VMs is associated with the creation of VMs and of their running on physical machines. In the context of cloud provisioning ``problems'', however, this article distinguishes between VM provisioning problem(s) and resource provisioning problem(s). Therefore, the problems associated with VM provisioning will decide the increase or decrease in the number of VMs and the problems associated with resource provisioning will determine the subsequent placement of the to-be-instantiated VMs on physical machines and the scaling of physical resources. The placement of VMs on physical machines is also known as ``VM placement'' \cite{2009_AutoVMManagement_NguyenVan, iqbal2010sla6} or ``VM packing'' \cite{2009_AutoVMManagement_NguyenVan}. Once, solutions to these problems have been generated, the decision-maker informs the provisioner of the required actions that need to be taken and then the actual provisioning occurs. 

In addition to above discussion, there is another reason for distinguishing between the two problems. When an AP decides to provision resources and makes an API call to instantiate a VM, it is the CP who decides on which particular physical machine would the instance reside, thus separating the decision making of the number of VM instances from their placement. Calheiros et al. \cite{calheirosvirtual3} and Quiroz et al. \cite{2009_TowardWorkloadProvCloud} also make the distinction between VM provisioning and resource provisioning. 

The following subsections discuss about provisioning problems and their association with application, VM and resource provisioning.

\subsection{Provisioning Problems}
\label{subsec:ProvProblems}
In this article, we consider dynamic provisioning to be associated with five \emph{problems}:
\begin{IEEEenumerate}
\item \textbf{Application scaling (\emph{AppScale})}: determining the increase and decrease in the number of applications or units that applications are composed of (e.g. by addition of replicas of processes or application tiers).
\item \textbf{Application placement (\emph{AppPlace}):} determining the placement of application units on VMs or on physical machines; the latter is for when VMs are not employed.
\item \textbf{VM Scaling (\emph{VmScale}):} determining the increase and decrease in the number of VMs and associated resources such as virtual CPUs, memory etc.
\item \textbf{VM placement (\emph{VmPlace}):} determining the placement of VMs on the machines, i.e. the allocation of resources to the VMs.
\item \textbf{Resource Scaling (\emph{ResScale}):} determining the increase and decrease in the number of operating resources, which the applications and/or the VMs will run on and utilize.
\end{IEEEenumerate}

\subsection{Provisioning types}
\label{subsec:ProvisioningTypes}
The different provisioning types are associated with the five problems identified above. Table~\ref{tab:DynProvAssocProblems} describes the relationship. As mentioned earlier, application and VM provisioning are the responsibility of AP, and CP is responsible for resource provisioning. If CP deploys the application then they adopt the role of the AP and therefore would control application and VM provisioning too.

\begin{table}[h]
\caption{Dynamic provisioning and associated Problems}
\label{tab:DynProvAssocProblems}
\centering
\begin{tabular}{|l|l|l|}
\hline
\heading{Provisioning} & \heading{Associated Problems} & \heading{Responsibility} \\
\hline

\parbox[t]{0.5cm}{Application\\ Provisioning \strut}
& \emph{AppScale} AND/OR \emph{AppPlace} 
& AP\\
\hline

\parbox[t]{0.5cm}{VM\\ Provisioning  \strut}
& \emph{VmScale} 
& AP\\
\hline

\parbox[t]{0.5cm}{Resource\\ Provisioning  \strut}
& \emph{VmPlace} AND/OR \emph{ResScale} 
& CP \\
\hline
\end{tabular}
\end{table}

\figurename~\ref{fig:CloudProv} describes in detail cloud provisioning from a problem solving perspective. The figure shows $n$ applications deployed on the cloud, where each application receives requests from $c$ client classes, where each client class expresses a distinct request pattern from a group of clients. Each application is composed of $t$ tasks that have a dependency structure between each other depending on the architecture of the application. The applications are run on $v$ VMs, which further run on $p$ physical machines. The mapping of the applications onto VMs solves the application provisioning problem. The tasks of the applications may have replicas and would be placed on the VMs. The determination of the number of VMs and its associated resources solves the VM provisioning problem. Finally, the resource provisioning solution places the VMs on various physical machines. As the workload from clients that the applications see fluctuate over time, the system manages and modifies the quantity and placement of resources accordingly. In this dynamic provisioning problem, solutions to the aforementioned problems would be used throughout the execution of the applications to determine any adjustments required in the system when receiving varying inputs.

\begin{table*}[!t]
\caption{Placement relationships and notations}
\label{tab:PlacementPolicies}
\centering
\begin{tabular}{|l|l|p{8cm}|}
\hline
\heading{Notation} & \heading{Relationship} & \heading{Description}\\
\hline

$A\xrightarrow{1..1}B$ & \emph{one-to-one}  & each A may be placed on one-and-only-one B \\
\hline

$C\xrightarrow{1..\infty}D$ & \emph{one-to-many} & each C may be placed on multiple D's, but multiple C's shall not reside within each D \\
\hline

$E\xrightarrow{\infty..1}F$ & \emph{many-to-one} & multiple E's may reside on one F, but these E's shall not span across multiple F's, except through replication of E\\
\hline

$G\xrightarrow{\infty..\infty}H$ & \emph{many-to-many} & multiple G's may reside on one H, and each G may span multiple H's \\
\hline
\end{tabular}
\end{table*}

\subsection{Horizontal and Vertical scaling}
\label{sec:HorizVertScaling}
Auto scaling is performed by either horizontal scaling or vertical scaling mechanisms. In horizontal scaling, a system is scaled through changes in the number of servers, e.g. replication of VM instances and addition of load balancers \cite{2011_DynamicallyScalingApps}. Likewise, vertical scaling denotes the changes to resources that are associated to an existing and running server, e.g. addition of processors to a live VM instance  \cite{2011_DynamicallyScalingApps}. Vertical scaling is made possible through ``hot plug'' \cite{2002_LinuxKernelHotplug}, a feature which allows for ``dynamic'' \cite{2010_DynamicProcDynamicOS} changes to the devices (e.g. CPU) connected to a running system, without a system shutdown. The Linux kernel supports the hot plug feature \cite{2002_LinuxKernelHotplug,2004_LinuxKernelHotplug}; however, Vaquero et al. \cite{2011_DynamicallyScalingApps} point out that changes owing to vertical scaling occur on live servers and majority of ``common operating systems'' \cite{2011_DynamicallyScalingApps} do not support manipulating the CPUs and memory available to the system without a system reboot.

In spite of the lack of support from various operating systems, as noted above, research has begun to focus on hot plug support (e.g. \cite{2010_DynamicProcDynamicOS, 2012_ChameleonOSDynamicProcs}) and on vertical scaling (e.g. \cite{2012_VerticalScaling}). Notably, VMware vSphere virtualization platform \cite{2014_VMwareVSphere} added a hot plug feature in version 4.0 known as ``hot add'' \cite{2009_VSphere4_HotAdd} for adding processors, memory and virtual disks to a VM instance, which is also present for version 5 of the platform \cite{2011_WhitePaper_VMwareVSphereCompRev}.

One advantage of vertical scaling over horizontal scaling as pointed by Yazdanov and Fetzer \cite{2012_VerticalScaling} is performance, where `VM instance acquisition time'' \cite{2012_VerticalScaling} is longer in horizontal scaling. As research advances in hot plug technology, more performance improvements would also be realized in vertical scaling. However, there are limits to how much a system can scale vertically, in which scenario horizontal scaling could be adopted, as done by Yazdanov and Fetzer \cite{2012_VerticalScaling} in their design.

\subsection{Resource relationships and placement policies}
\label{subsec:PlacPolicies}
The placement of applications on VMs and the placement of VMs on machines are governed by their placement policies, which are described by the relationships that exist between the resources or entities. In this article, we introduce general relationships and notations that help in expressing these placement policies. It is important to note that defining relationships and having a placement policy affects how the provisioning algorithm works and how the resources are scaled, although, defining a placement policy does not necessarily mean a placement algorithm is proposed.

Table~\ref{tab:PlacementPolicies} lists the relationships and their notations. In these relations, the entity on the left-side is placed onto the entity on the right-side. To express application placements, the entities on the left may be substituted by either application, tier or task, whereas in context of VM placement, the left-side entities would be VM. Following this, the right-side entity would be VM for application placement and host machine for VM placement. In the table, $A\xrightarrow{1..1}B$, describes a  direct \emph{one-to-one} relation between A and B, and that each A may be placed on one-and-only-one B, e.g. each application is allocated it's own VM or in case of VM placement, each VM would run on a separate host machine. In this relationship the entities that are placed may be replicated, however, a one-to-one mapping is still followed. $C\xrightarrow{1..\infty}D$, describes a \emph{one-to-many} relation between C and D, where each C may be placed on multiple D's, but multiple C's shall not reside within each D, e.g. a dedicated hosting scenario where the components of an application --- such as tasks or tiers --- are placed across many VMs, but the VMs are not shared among different applications, and each VM would thereby include components of one application only. $E\xrightarrow{\infty..1}F$, describes a \emph{many-to-one} relation between E and F, where multiple E's may reside on one F, but these E's shall not span across multiple F's, except by replication of E, e.g. a VM may host various applications but each of these applications are contained within the VM and their components are not spread across other VMs, unless the application is replicated as a whole. $G\xrightarrow{\infty..\infty}H$, describes a \emph{many-to-many} relation between G and H, where multiple G's may reside on one H, and each G may span multiple H's, e.g. a shared hosting scenario where each application is spread across multiple VMs, and each VM is hosting multiple different applications. These relationships can be used to describe complex policies and these notations can be extended further to accommodate more specific relationships or restrictions that may exist.

\section{When is the provisioning decision made?}
\label{sec:ReactiveProactiveProv}
Proactive provisioning is made possible through use of workload predictors, performance models and system monitors (e.g.  \cite{calheirosvirtual3}). Through workload prediction, the future incoming workload can be predicted and this would feed as input to the performance model. The parameters of the performance model would be determined through monitoring the system and through prediction modules \cite{calheirosvirtual3}. The  performance metrics thus obtained from solving the model would indicate if QoS objectives would be satisfied through the current system configuration. If the objectives are not met then the system configuration is tweaked through the performance model, following an iterative process until the objectives are satisfied or there is an indication that the objectives cannot be met. The system configuration that led to meeting of the objectives by solving the performance model is then applied to the real system such that the system can handle the incoming workload. In contrast to proactive provisioning, reactive provisioning proceeds with modification of system configuration as the workload changes. Example of reactive provisioning includes work by Iqbal et al. \cite{iqbal2010sla6}, who discuss a provisioning prototype implementation in the Eucalyptus Cloud, which horizontally scales the system based on the bottleneck tier found. Amazon auto scaling \cite{AmazonAutoScaling} provided by Amazon web services also allow for reactive addition/removal of VM instances in case of increase/decrease seen in the CPU resource utilizations. Furthermore, this auto scaling feature also allows for specifying a preprogrammed schedule of the scale.

Urgaonkar et al. \cite{urgaonkar2008agile10} propose both reactive and proactive/predictive provisioning approach for their provisioning algorithm. Although, their work is not applied to cloud systems, their approach is still applicable to the cloud provisioning domain. With proactive technique the dynamic changes to the system configuration are based on workload prediction from previously seen long-term workload demands. The reactive technique complements the previous approach by adjusting the configuration when sudden burst in workload is seen occurring within a short time duration. Zhang et al. \cite{zhang2010cloud13} have also emphasized the importance and necessity of both reactive and proactive methods.

\section{What resources are provisioned?}
\label{sec:ServiceProv}
Various resources are provided by clouds as services:
\begin{IEEEenumerate}
\item VMs and associated resources are provisioned through IaaS
\item Development platform is made available through PaaS
\item Software is accessible through SaaS
\end{IEEEenumerate}

The previous sections and in particular, Section~\ref{sec:TypesCloudProvTerm} and \ref{sec:TowardPerCloudProv} provide details about provisioning of VMs and software applications. In this section, we mention about provisioning of platform and other resources. 

Vaquero et al. \cite{2011_DynamicallyScalingApps} discuss about scaling of both servers and platforms in their paper with focus toward applications. They mention that rather than scaling based on each VM only, the other approach would be a controller that manages resources at a higher level of abstraction, more specifically by considering a complete application, thereby these application level conditions would be mapped to the abstraction made available by the cloud through their API (e.g. VM scaling). With regards to scaling of platforms by the PaaS providers, the authors discuss about containers and databases, the main entities of the platform. For scaling containers, component state has to be taken into consideration, and approaches may include state-aware components, replication of state information or caching. For databases, the following approaches may be adopted: ``distributed caching, NoSQL databases, and database clustering'' \cite{2011_DynamicallyScalingApps}.

Alongside, other resources such as physical servers \cite{2013_MAAS_Canonical,2012_MAASEffect} and network \cite{2012_NAAS_Costa,2011_DynamicallyScalingApps} may be also be provisioned depending on them being made available as services by CP and on state-of-art of research in these fields. ``Metal as a service'' (MAAS) \cite{2013_MAAS_Canonical} --- which allows for provisioning of physical server nodes directly instead of VMs --- is already offered by Canonical Inc. either by a package installation on Ubuntu operating system or through a newly installed Ubuntu Server. With regards to network, Vaquero et al. \cite{2011_DynamicallyScalingApps} discuss about the importance of scaling networks, pointing that in ``consolidated datacenter scenario''  \cite[p.~45]{2011_DynamicallyScalingApps} because many VMs use a shared network, a lot of network bandwidth may eventually be required. The usual approach adopted in this scenario is toward resource over-provisioning to meet (network) demands, which the paper argues against, as the method shows a lack of consideration toward applications that do not always consume the complete network bandwidth (alloted), hinting that such an arrangement leaves the network resources under-utilized. The authors suggest toward solutions that take ``actual network usage'' \cite[p.~47]{2011_DynamicallyScalingApps} into consideration, either by measuring network used by each application or by having the applications themselves ask for more network bandwidth. 

Just like with interest in offering of network and physical servers, as cloud computing use grows there will be other resources whose importance may come to be realized by users, which would warrant further research and those resources would eventually come to be delivered as cloud services.

\section{How are the provisioning problems solved?}
\label{sec:ProvAlgorithms}
In the following subsections we mention the algorithms, techniques and technologies that facilitate the provisioning in cloud systems.

\subsection{Algorithms}
\label{subsec:algorithms}
Various algorithms have been suggested and applied to solve cloud dynamic provisioning problems and these are mentioned below. Although a few of these have been proposed before the advent of cloud computing they are still very applicable to solving provisioning in the cloud domain.

Hill climbing along with performance models have been used by Menasce et al. \cite{menasce2003automatic8} for finding optimal system configuration parameters. In dynamic provisioning, hill climbing algorithm has been applied by Zheng \cite{zheng2007model9} for dynamic provisioning of web and database servers, for admission control and thread count management. Buyya et al. \cite{2010_Buyya_InterCloud}, in relation to finding solutions to the differing QoS and optimization goals of cloud services, mention that for such cases we have a multi-dimensional optimization problem. To solve such problems, ``one can explore multiple heterogeneous optimization algorithms, such as dynamic programming, hill climbing, parallel swarm optimization, and multi-objective genetic algorithm'' \cite[p.~22]{2010_Buyya_InterCloud}. Greedy algorithms, genetic algorithm, and various vector packing algorithms have been applied by Stillwell et al. \cite{2010_ResourceAllocationAlgorithms} to solve the resource allocation problem in virtualized platforms. Simulated annealing algorithm has been proposed by Pandit et al. \cite{2014_ResourceAllocSimAnnealing} for solving resource allocation in the clouds. Furthermore, Pandit et al. \cite{2014_ResourceAllocSimAnnealing} has modeled the problem as a variant of multi-dimensional bin packing problem. The relationship between APP with bin-packing and multiple knapsack problems has explained by Urgaonkar et al. \cite{2007_Urgaonkar_APP}. They propose various approximation algorithms and heuristics such as First-Fit, Max-First, Best-Fit, Worst-Fit, etc. to solve the APP. As part of provisioning, for load prediction various algorithms have also been proposed. Xiao et al. \cite{2013_Xiao_DynResAllocationVM} have presented their own algorithm: ``Fast Up and Slow Down (FUSD) algorithm'' \cite{2013_Xiao_DynResAllocationVM}, that predicts the expected resource utilization and helps toward making more stable provisioning judgements. Alongside, they have also looked at linear autoregression models for prediction and provide a comparison with their algorithm. Other techniques for load prediction include using artificial neural networks (e.g. Chabaa et al. \cite{2010_IdentPredictInternetTraffic}, Prevost et al. \cite{2011_PredictionCloudNN}).

\subsection{Techniques and Technologies}
\label{subsec:techniquesTechnologies}
Provisioning in clouds is made possible through support of various techniques and technologies such as  hot-plugging, VM migration, VM resizing, performance modeling, workload prediction, admission control, system monitoring, server consolidation and load balancing. As discussed earlier in section \ref{sec:HorizVertScaling}, the hot-plugging technology makes vertical scaling possible. There are also interesting mechanisms such as: VM migration, VM sizing and server consolidation \cite{2012_Mishra_DynamicResourceManagement} that play a key role in provisioning and used mainly for CP-based resource provisioning. Xiao et al. \cite{2013_Xiao_DynResAllocationVM} have relied on VM migration and server consolidation technologies in their dynamic provisioning approach. Calcavecchia et al. \cite{2012_VMPlacementStratCloud} have used VM migrations for load balancing between the hosts machines. Performance modeling to help make dynamic provisioning decisions have been employed by many including Li et al. \cite{jimliwoodside2009performance5}, Huber et al. \cite{huber2011model4}, Shoaib and Das \cite{2012_ucc2012YasDas}, and Li et al. \cite{2011_JimLiCloudOpt}. Calheiros et al. \cite{calheirosvirtual3} in their solution have employed multiple techniques such as workload prediction, performance modeling and VM monitoring. Multiple techniques such as admission control, VM provisioning, multiple job queues, request priority and performance monitoring have been employed by Das et al. \cite{2013_AKDas_IntelligentApproachVMProvisioning} to meet response times of requests sent to a cloud.

\section{Toward performance-oriented cloud provisioning}
\label{sec:TowardPerCloudProv}

Performance of computer hardware and software systems has been the focus of many works such as that those by Lazowska et al. \cite{1984_LazowskaQSP_ref29}, Smith and Williams \cite{2002_PerfSolutions_ref23}, and Menasce et al. \cite{2001_Menasce_CapacityPlanning, 2004_Menasce_PerformanceByDesign}. These works have explained performance and described the performance modeling process. They have also emphasized the importance of considering performance within the system development process such that performance objectives are satisfied when the system is ready. Such performance objectives may be ``response time, throughput, or constraints on resource usage'' \cite[p.~29]{2002_PerfSolutions_ref23}. Research in performance of web systems has been conducted by many researchers including Dilley et al. \cite{1998_Dilley_Ref22}, Menasce et al. \cite{2001_Menasce_CapacityPlanning}, Liu et al. \cite{2004_LiuPerfEngineeringJava_ref4}, Ufimtsev and Murphy \cite{2006_Ufimtsev_ref6} and Urgaonkar et al. \cite{2007_UrgaonkarAnalyticModeling_ref8}. Menasce \cite{menasce2003automatic8} in their dynamic system reconfiguration approach present a QoS controller design that uses hill-climbing algorithm, relying on monitored data and QN models to derive an optimal system configuration (e.g. thread count and maximum queue size) for meeting the objectives of response time, throughput and rejection probability of a multi-tiered electronic commerce website. Gradually, the focus of research moved toward dynamic provisioning, which previous works have dedicatedly looked into, and these efforts have been extended to find their application in the cloud computing domain as well. The important works below highlight these efforts and show the gradual progression toward performance-oriented cloud provisioning.

Karve et al. \cite{2006_KarveTantawi_DynamicPlacementClusteredWebApps} describe the design of a middleware platform and controller, where the number of application instances are dynamically adjusted and placed on machines as per the demands of the application. This work adds to their previous work \cite{2005_KimbrelTantawi_DynamicPlacementService}, by achieving better balance of application load on the machines and providing improvements to their algorithm by minimizing number of placement changes --- through the use of ``incremental placement''\cite{2006_KarveTantawi_DynamicPlacementClusteredWebApps} approach ---  and maximizing the achieved application demand. In their work they consider CPU and memory resource capacities of the machines and the respective requirements of the applications in making placement decisions.

Urgaonkar et al. \cite{2007_Urgaonkar_APP} provide the algorithms that solve the application placement problem (APP) for applications running on clusters and show that these problems are NP-hard. They describe how APP relate to bin-packing and multiple knapsack problems. Online and Offline APP as two types of placement problems that are discussed. Offline APP compute placements without taking the order of the applications into consideration, whereas online APP compute placements one-by-one for each application, placing applications from lower to higher indices, and not allowing changes in placement of applications that were placed earlier than the application that is currently being considered for placement. Depending on the variants of online and offline APP, as per the different placement restrictions, they propose various approximation algorithms and heuristics such as First-Fit, Max-First, Best-Fit, Worst-Fit, etc. to solve the APP.

Zheng \cite{zheng2007model9} presents a framework that automatically allocates application and database server resources when response time objectives are violated. Their controller uses hill-climbing algorithm, which begins at a initial system state and evaluates the neighbor states, choosing the neighbor that results in the lowest cost. In a similar pattern, the algorithm continues searching through the neighbors until the optimal cost is found. Through a few simulation experiments and a case study, the workings of the provisioning framework and the controller are demonstrated. The case study employs resource provisioning where the needed servers are added to the application server cluster from a pool of available servers. Including dynamic provisioning, admission control and thread count management are also employed.

Urgaonkar et al. \cite{urgaonkar2008agile10} propose a combination of workload prediction, virtualization technology, admission control, proactive and reactive provisioning to meet given response time deadlines of multi-tier systems. The QN model receives inputs from workload prediction to find ``the number of servers to be allocated to each tier based on the estimated workload'' \cite[p~1:5]{urgaonkar2008agile10}. Use of VMs help in quick adjustment of resources to the application tiers. In this particular work they consider a ``dedicated hosting'' platform \cite{urgaonkar2008agile10}, where each server runs one application and each application may spread across multiple servers. This type of hosting is different from previous works on ``shared hosting'' platforms \cite{2009_Urgaonkar_ResourceOverbookingShared, 2004_Urgaonkar_SharcShared}, where applications would not only spread across multiple servers but also share server resources, thereby allowing multiple applications to run on a server. By proposing and demonstrating the use of VMs for quick resource adjustments for multi-tier applications, this work \cite{urgaonkar2008agile10} laid a strong foundation for future research work on provisioning in the clouds and is one of most detailed and significant contributions before research in cloud provisioning began in earnest. 

Iqbal et al. \cite{iqbal2010sla6} discuss their application and VM provisioning prototype implementation, which they have extended from their earlier work \cite{2009_Iqbal_OneTier} on one-tier to two-tier web applications, in the Eucalyptus Cloud. First, the bottleneck tier is found using a simple algorithm based on CPU utilizations and response times, and then VMs are added such that response times guarantees are not violated. They further plan to improve their algorithm to include n-tier applications. 

Kijsipongse et al. \cite{kijsipongse2010autonomic25} propose an architecture which dynamically provisions VMs from a remote cloud according to one of the two provisioning policies introduced by the authors and adds the VM as part of a local cluster (management) system. A test-bed is setup where Eucalyptus cloud is used, which can host a maximum of six VMs on the cloud. The focus of this work is on the job scheduling and VM provisioning policies with related affect on job queue size.

Li et al. \cite{jimliwoodside2009performance5} use Layered Queueing Network (LQN) performance models \cite{2009_Franks_EnhancedModelingLQN}, which are analytical models based on extended QN, along with network flow model (NFM) to solve the deployment optimization problem in maximizing the profit of the cloud while considering response time or throughput as constraints. Their combined application, VM and resource provisioning approach meets objectives by optimally placing VMs on physical machines, where each application task is hosted on one VM.

Van et al. \cite{2009_AutoVMManagement_NguyenVan} in their dynamic provisioning approach differentiate VM provisioning from VM placement and handle them in separate steps. It is key to note that their definition of VM provisioning also incorporates finding placements for the applications on the VMs (i.e. application placement problem), which based on our definition is associated with application provisioning. Irrespective of this, their approach deals with solving the cloud-based provisioning problems considering both performance and cost, where these problems as expressed as constraint satisfaction problems. Techniques for facilitating provisioning include instantiation and destruction of VMs, VM migration, and VM horizontal and vertical scaling; where vertical scaling is achieved through resizing the VMs.

Huber et al. \cite{huber2011model4} present a framework allowing dynamic allocation and de-allocation of VMs and virtual CPUs, based on the run-time demands of the virtualized systems. The framework comprises of a ``control loop'' \cite{huber2011model4} and uses performance models. A case study using a SPECjEnterprise2010 benchmark deployment was conducted for three scenarios where a new service was added, workload was increased and decreased. The evaluation showed promising results with meeting response time guarantees by dynamically managing resources. For an example week workload, the paper showed maximum savings of 40\% of resources may be achieved when using the dynamic approach in comparison to a static allocation that would meet QoS guarantees.

Apart from previous works, Calheiros et al. \cite{calheirosvirtual3} present a unique perspective where VM and application provisioning is done by those providing both SaaS and PaaS (i.e. AP), whereas resource provisioning is reserved for IaaS providers, mainly due to the lack of ``control'' \cite{calheirosvirtual3} regarding VM placement available to the AP. Similar idea of ``decoupled control'' \cite{2009_AutomatedControlCloudChallenges} has previously been suggested by Lim et al. 

In their proposed algorithm, Calheiros et al. \cite{calheirosvirtual3}, aim to satisfy QoS requirements such as response time, utilization and rejection rate of VMs based on negotiated QoS attributes. The solution employs workload prediction and performance modeling and uses VM monitoring. A simple QN model of the system is depicted. Furthermore, admission control is adopted by determining the maximum queue size of a queueing station and rejecting requests that arrive if the queue is full. The simulation results obtained by using CloudSim \cite{2011_CloudSimAToolkit}, for a 1000 host data center, show promising results where the VM hours were reduced --- with no or about none rejection rate --- and the negotiated response times were met.

Chi et al. \cite{chiheuristic28} provide a heuristic search algorithm to find optimal deployment configuration for a multi-tiered web application in the clouds. A given configuration includes the number of VMs allocated to each tier. The main features of the algorithm include a utility function that adds another level of QoS along with response time, a rule-set database to find initial configuration for a given workload, and  a pruning algorithm which finds the optimal deployment. The pruning algorithm adds/removes VMs of tiers of an application and uses differences in workloads and the utility function to reach optimal solution quicker. Alongside, performance models are used for decision making. They evaluate their approach with a two-tier web application. For finding optimal configuration they use an expert system --- relying on and updating the rule-set database --- and their pruning algorithm.

Li et al. \cite{2011_JimLiCloudOpt} have developed ``CloudOpt'' \cite{2011_JimLiCloudOpt}, an approach that relies on various methods such as network flow models (NFM),  performance models and  bin packing heuristics for solving problems associated with application, VM and resource provisioning in the clouds. The optimization problem is formulated as a mixed integer problem, which is derived from the NFM, and the performance models are used to include resource contentions. They consider a mapping between an application task and a VM; however, application and VM scaling together is achieved by creating replicas of tasks, whereas application and VM placement is achieved by moving those replicas between hosts. Their approach is applicable to dynamic provisioning environments when the CloutOpt optimization is ``carried out periodically'' \cite{2011_JimLiCloudOpt}. The focus of their work is on meeting response time objectives and minimizing cost, while also considering software licence and memory constraints.

\begin{table*}
\caption{Placement policies adopted by various publications in dynamic provisioning}
\label{tab:PlacementPoliciesRelatedWorks}
\centering
\begin{tabular}{|l|l|c|c|}
\hline
\heading{Year} & \heading{Reference} & \heading{\parbox[t]{2.5cm}{\centering{Cloud\\or\\Virtualization based\vspace{0.5em}}}} & \heading{Provisioning Policy} \\
\hline


2006 & 
Karve et al. \cite{2006_KarveTantawi_DynamicPlacementClusteredWebApps} & 
No & 
\parbox[t]{10cm}{\centering application~$\xrightarrow{\infty..\infty}$~host \strut} \\ 
\hline

2007 & 
Urgaonkar et al. \cite{2007_Urgaonkar_APP}& 
No & 
\parbox[t]{10cm}{\centering Various policies are considered here, e.g. many-to-many relationship between application components and hosts, and one-to-one relationship between applications and hosts, etc. \strut}  \\ 
\hline

2007 & 
Zheng \cite{zheng2007model9} & 
No & 
\parbox[t]{10cm}{\centering application or tier~$\xrightarrow{1..1}$~host \strut} \\ 
\hline

2008 & 
Urgaonkar et al. \cite{urgaonkar2008agile10} & 
Yes (Virtualization) & 
\parbox[t]{10cm}{\centering tier~$\xrightarrow{1..1}$~VM\\ VM~$\xrightarrow{1..1}$~host \\ Although multiple VMs reside and run on a host, only one VM is active at a time, therefore a one-to-one relationship is assumed to exist between VMs and hosts in this case. \\~\\ application~$\xrightarrow{1..\infty}$~host\\ Indirectly, each application may be placed on many hosts. \strut} \\
\hline

2010 & 
Iqbal et al. \cite{iqbal2010sla6}  & 
Yes & 
\parbox[t]{10cm}{\centering tier~$\xrightarrow{1..1}$~VM \\ application~$\xrightarrow{1..\infty}$~VM\strut} \\ 
\hline

2010 & 
Kijsipongse et al. \cite{kijsipongse2010autonomic25} & 
Yes & 
\parbox[t]{10cm}{\centering job-type~$\xrightarrow{1..1}$~VM \strut}\\ 
\hline

2009 & 
Li et al. \cite{jimliwoodside2009performance5} & 
Yes & 
\parbox[t]{10cm}{\centering task~$\xrightarrow{1..1}$~VM \\ VM (or task)~$\xrightarrow{\infty..1}$~host \strut}\\ 
\hline

2009 & 
Van et al. \cite{2009_AutoVMManagement_NguyenVan} & 
Yes & 
\parbox[t]{10cm}{\centering application~$\xrightarrow{1..\infty}$~VM \strut} \\ 
\hline

2011 & 
Huber et al. \cite{huber2011model4} & 
Yes & 
\parbox[t]{10cm}{\centering service~$\xrightarrow{\infty..1}$~VM \strut} \\ 
\hline

2011 & 
Calheiros et al. \cite{calheirosvirtual3} & 
Yes & 
\parbox[t]{10cm}{\centering application instance~$\xrightarrow{1..1}$~VM \\ an application instance is characterized as ``software library, executable, data, or functional component'' \cite[p.297]{calheirosvirtual3} \strut} \\ 
\hline

2011 & 
Chi et al. \cite{chiheuristic28} & 
Yes & 
\parbox[t]{10cm}{\centering tier~$\xrightarrow{1..1}$~VM \\ application~$\xrightarrow{1..\infty}$~VM \strut} \\ 
\hline

2012 & 
Calcavecchia et al. \cite{2012_VMPlacementStratCloud} & 
Yes & 
\parbox[t]{10cm}{\centering VM~$\xrightarrow{\infty..1}$~host \strut} \\ 
\hline

2013 & 
Casalicchio et al. \cite{2013_Casalicchio_AutoResProv} & 
Yes & 
\parbox[t]{10cm}{\centering VM~$\xrightarrow{\infty..1}$~host \strut}\\ 
\hline

2013 & 
Xiao et al. \cite{2013_Xiao_DynResAllocationVM} & 
Yes & 
\parbox[t]{10cm}{\centering application~$\xrightarrow{\infty..1}$~VM \\ VM~$\xrightarrow{\infty..1}$~host \strut}\\ 
\hline

2011 & 
Li et al. \cite{2011_JimLiCloudOpt} & 
Yes & 
\parbox[t]{10cm}{\centering task~$\xrightarrow{1..1}$~VM \\  VM (or task)~$\xrightarrow{\infty..1}$~host \strut}\\ 
\hline

2012 & 
Shoaib and Das \cite{2012_ucc2012YasDas} & 
Yes & 
\parbox[t]{10cm}{\centering task~$\xrightarrow{\infty..1}$~VM \\ application~$\xrightarrow{\infty..\infty}$~VM \strut} \\ 
\hline

2013 & 
Das et al. \cite{2013_AKDas_IntelligentApproachVMProvisioning} & 
Yes & 
\parbox[t]{10cm}{\centering job-type~$\xrightarrow{1..1}$~VM \strut}\\ 
\hline

2010 & 
Stillwell et al. \cite{2010_ResourceAllocationAlgorithms} & 
Yes (Virtualization) & 
\parbox[t]{10cm}{\centering VM~$\xrightarrow{\infty..1}$~host\\~\\ Two application placement policies are discussed:\\ (i) service~$\xrightarrow{1..1}$~VM \\ (ii) service~$\xrightarrow{1..\infty}$~VM \\ tier~$\xrightarrow{1..1}$~VM \strut}\\
\hline

2014 & 
Pandit et al. \cite{2014_ResourceAllocSimAnnealing} & 
Yes & 
\parbox[t]{10cm}{\centering request~$\xrightarrow{\infty..1}$~resource \\ VM~$\xrightarrow{\infty..1}$~host \strut}\\ 
\hline

\end{tabular}
\end{table*}

Shoaib and Das \cite{2012_ucc2012YasDas} present a detailed provisioning algorithm that adds VMs or virtual processors based on software and hardware bottlenecks identified, while considering various limits, such that the performance goals specified and met. The results from a case study show that using of simple bottleneck detection approach based on utilization of processors in comparison to the ``layered bottlenecks'' \cite{franks2006layeredBottlenecks26} approach leads to use of more resources to meet performance objectives.

Das et al. \cite{2013_AKDas_IntelligentApproachVMProvisioning} meet response times of requests sent to a cloud through employment of admission control, VM provisioning, multiple job queues, request priority and performance monitoring. They use simulation performed using CloudSim \cite{2011_CloudSimAToolkit} to show their provisioning approach performs better than a simple VM provisioning approach, by showing results on admitted/served requests, rejected requests and VM instantiation time.

Calcavecchia et al. \cite{2012_VMPlacementStratCloud} present ``Backward Speculative Placement (BSP)'' \cite{2012_VMPlacementStratCloud}, a technique that processes VM deployment requests received by a CP to optimize the placements of VMs on heterogeneous physical machines. The objectives considered are meeting of the CPU demands of the VMs, minimizing the VM migrations and balancing of the load between the hosts. The technique uses two phases for the decisions, where one phase handles new requests for VM deployment and another phase periodically optimizes the existing placement (through VM migrations). Through simulation results they show that their approach generates placements that meet ``high level of demand satisfaction'' \cite{2012_VMPlacementStratCloud}. Owing to VM migrations, the number of physical machines that are active changes in their problem, and therefore we consider their approach to also includes solving of the resource scaling problem along with the VM placement problem.

Casalicchio et al. \cite{2013_Casalicchio_AutoResProv} also focus on the VM placement problem faced by CP. The objective is maximize the revenue of the CP while meeting resource, availability and VM migration constraints. Their approach presents an algorithm based on hill climbing to solve the problem.

Xiao et al. \cite{2013_Xiao_DynResAllocationVM} in their article explain their approach to dynamic VM placement and resource scaling in the clouds. The aim is toward placing VMs on machine such that the requirements of VMs are met and also toward minimizing the number of machines used. Through VM migrations the overload on machines is decreased and idle machines are turned off (or set to standby). Their approach is based on a load prediction algorithm (``Fast Up and Slow Down (FUSD) algorithm'' \cite{2013_Xiao_DynResAllocationVM}) and a ``skewness'' \cite{2013_Xiao_DynResAllocationVM} metric; the former is used for the prediction of expected resource utilization, and the latter is related to the utilization of multiple resources on each machine and is minimized in the algorithm. They present results from both simulation and measurements to demonstrate the applicability of their approach.

Pandit et al. \cite{2014_ResourceAllocSimAnnealing} have proposed a simulated annealing algorithm for resource allocation in the clouds and model this problem as a variant of multi-dimensional bin packing problem. They explain their bin packing problem modeling through a simple VM placement example. Through simulation they show that their proposed algorithm performs better than First-Come-First-Serve algorithm by having a higher average resource utilization when mapping the requests to the resources. We consider that their approach relates to resource allocation problems in general, such as application placement and VM placement.

Stillwell et al. \cite{2010_ResourceAllocationAlgorithms} have proposed algorithms for resource allocation in virtualized shared platforms running homogeneous machines. Although their approach focuses on static workloads, by periodically finding the allocations, their approach would be applicable for dynamic workloads in dynamic provisioning scenarios. They formulate the problem of resource allocation as a mixed integer linear program (MILP), where the objective is to maximize performance and fairness through a metric known as ``minimum yield'' \cite{2010_ResourceAllocationAlgorithms}. The article is mostly focused on allocation of services to machines, where each service runs within one VM, although services that run on multiple VMs are also included in the discussion. The proposed algorithms include greedy algorithms, genetic algorithm, and vector packing algorithms (multi-dimensional bin packing). Since the article focuses on resource allocation problem in general, it is applicable to solving both VM placement and application placement problems.

Alongside with VM provisioning, researchers have also considered how applications relate to VMs. Calheiros et al. \cite{calheirosvirtual3} in their VM and application provisioning approach as discussed above, have considered a ``one-to-one mapping relationship between an application instance \ldots\ and a VM instance'' \cite{calheirosvirtual3}. Van et al. \cite{2009_AutoVMManagement_NguyenVan} in their work mention that an application can be associated to many VMs, and each VM is related to one application, which is similar to ``dedicated hosting'' \cite{urgaonkar2008agile10}. The relation between applications and VMs is important one to determine when dealing with application placement problems. Table~\ref{tab:PlacementPoliciesRelatedWorks} lists research publications in dynamic provisioning that have been mentioned in this section, and mentions the placement policies that have been adopted by them.

Including performance, dynamic provisioning could be employed to achieve various goals. Researchers have used their approach to meet multiple objectives such as minimizing cost and meeting performance objectives (e.g. \cite{jimliwoodside2009performance5, 2009_AutoVMManagement_NguyenVan, 2012_VerticalScaling}), or addressing availability and performance (e.g.~\cite{2010_Jung_PerfAvailability}), or considering energy and performance~(e.g.~\cite{2012_Anton_EnergyAware}).

\section{Conclusions}
\label{sec:Conclusions}
This survey article begins with explaining cloud provisioning and its different types, alongside clarifying the terminology used to describe them. A detailed classification of cloud dynamic provisioning is presented next, followed by mentioning of key related research work that contributed to dynamic provisioning approaches in general and recent research efforts in application, VM and resource provisioning. Along with the classification, this article contributes by explaining the different facets of provisioning in the clouds and in particular explaining in detail the following: reactive and proactive provisioning, horizontal and vertical provisioning, provisioning of resources as services, provisioning goals, algorithms, techniques and technologies involved.

We make the following observations through our survey. First, cloud provisioning is used for various purposes including but not limited to following: minimizing resource usage, minimizing cost, meeting QoS, achieving multiple objectives, where algorithms should be quick, cause minimal changes to the existing allocation scheme, and handle different resource types. Second, not all publications explicitly or clearly mention the provisioning and placement policies that they use in their work. This survey introduces notations that can serve as a very useful means for explaining and identifying the policies used by a provisioning scheme.


%



\ifCLASSOPTIONcompsoc
  \section*{Acknowledgements}
\else
  \section*{Acknowledgement}
\fi

A small portion of this work has been presented earlier in the \emph{Proceedings of the 2012 IEEE/ACM Fifth International Conference on Utility and Cloud Computing (UCC 2012)}. We would like to thank NSERC for their support.

\ifCLASSOPTIONcaptionsoff
  \newpage
\fi



%
\bibliographystyle{IEEEtran}
\bibliography{CloudProvisioning}

%

\begin{IEEEbiography}{Yasir Shoaib}
Biography text here.
\end{IEEEbiography}

\begin{IEEEbiography}{Olivia Das}
Biography text here.
\end{IEEEbiography}






\end{document}